\documentclass[runningheads]{llncs}
\usepackage[utf8]{inputenc}
\usepackage{hyperref}
\usepackage{xcolor}
\usepackage{subcaption}
\usepackage{graphicx}
\usepackage{makecell}
\usepackage{amsmath}
\usepackage{booktabs}
\newcommand{\CV}{\mathcal{V}}
\newcommand{\CC}{\mathcal{C}}
\newcommand{\CL}{\mathcal{L}}
\newcommand{\ria}{\rightarrow}
\newcommand{\set}[1]{\{\,#1\,\}}

\DeclareMathOperator*{\argmax}{arg\,max}

\begin{document}

\title{Welfare effects of strategic voting under scoring rules\thanks{Support from the Basic Research Program of the HSE University is gratefully acknowledged.}}

\author{Egor Ianovski\inst{1}\orcidID{0000-0001-9411-2529} \and
Daria Teplova\inst{2}\orcidID{0000-0003-0511-6507} \and
Valeriia Kuka\inst{3}}

\authorrunning{E. Ianovski et al.}

\institute{HSE University, St Petersburg, Russia\\
\email{george.ianovski@gmail.com} \and
ITMO University, St Petersburg, Russia
 \and
Saint Petersburg State University, St Petersburg, Russia}
\maketitle             

\begin{abstract}
Strategic voting, or manipulation, is the process by which a voter misrepresents his preferences in an attempt to elect an outcome that he considers preferable to the outcome under sincere voting. It is generally agreed that manipulation is a negative feature of elections, and much effort has been spent on gauging the vulnerability of voting rules to manipulation. However, the question of why manipulation is actually bad is less commonly asked. One way to measure the effect of manipulation on an outcome is by comparing a numeric measure of social welfare under sincere behaviour to that in the presence of a manipulator. In this paper we conduct numeric experiments to assess the effects of manipulation on social welfare under scoring rules. We find that manipulation is usually negative, and in most cases the optimum rule with a manipulator is different to the one with sincere voters.

\end{abstract}

\section{Introduction}

In the parlance of social choice, voting is the mechanism by which a group of agents aggregate their preferences over a set of outcomes to select a single outcome for the whole of society. The language naturally brings to mind a political election -- and indeed, modern voting theory traces back to the proposals of electoral reform by Borda and Condorcet in 18th century France -- but the model can equally well describe the process by which a panel of judges determines the winner of a contest, a firm makes its hiring decisions, or an ensemble of algorithms comes to a joint decision.

Strategic voting is the process by which an agent casts a vote not in agreement with his true preferences, but in an attempt to attain the most preferable outcome -- the standard example is a supporter of a minor third party voting for a major party closest to his ideological position. Strategic voting is also known as manipulation, a much more coloured term that is more readily associated with fraud, bribery, and other malfeasance, rather than the innocent act of not wasting one's vote. In a political context, one could even go so far as to claim it is the democratic duty of every citizen to ``manipulate'' in this fashion to make his voice heard. To understand why manipulation is perceived negatively, one needs to consider the other applications of voting, those involving a small panel of experts choosing among outcomes, particularly when some outcomes are objectively ``better'' or ``more correct'' than others. In the case of a sporting event, we would certainly hope the judges rank the athletes based on their objective performance rather than personal preference (though we understand well enough that this is not always the case \cite{Zitzewitz2006}).

The Gibbard-Satterthwaite theorem \cite{gibbard,satterthwaite1975strategy} established that all non-trivial voting rules are vulnerable to manipulation, but that does not mean that all rules are equally vulnerable, and much attention has been focused determining that degree of vulnerability by measuring the likelihood of manipulation taking place \cite{nitzan1985vulnerability,kelly1993almost,aleskerov2015manipulability}, the complexity of finding such a manipulation \cite{bartholdi1989computational,Conitzer_2007,Walsh2011}, or the amount of information necessary to manipulate \cite{chopra2004knowledge,xia2011determining,conitzer2011dominating}.

While these approaches greatly improved out understanding of how hard or how likely it is to affect the outcome of a voting rule, they said nothing about whether or not this manipulation is actually bad. In a sense, they did not have the vocabulary to do so -- the twentieth century was dominated by the axiomatic approach to voting, studying which properties a voting rule does or does not satisfy, and the question of why we vote (presumably, to elect good outcomes) was rarely asked.

This was not always the case. Condorcet motivated his method as electing the objectively ``correct'' candidate \cite{Young1995}, while Borda's contemporaries observed that his method can be seen to maximise voter satisfaction \cite{Laplace1795}. This welfare approach to voting has resurfaced in recent years, and suggests a clear means to measure the effect of manipulation on voting rules -- the degree to which the final social welfare of society is affected.

In this paper we numerically explore the welfare effects of strategic voting under scoring rules, focusing on the case of a small number of voters making a choice between candidates, and the final outcome being evaluated in terms of Borda, Rawls, or Nash welfare.

\subsection{Related work}

To our knowledge, the first authors to study the welfare effects of strategic voting were Chen and Yang \cite{Chen2002}. In their work they consider voting in open primaries, where members of party $B$ are allowed to vote for the candidate party $A$ will nominate for the general election. Such primaries invite strategic behaviour in the form of ``mischief voting'' -- voters of party $B$ will strive to nominate the most extreme candidate of party $A$, to guarantee that party $B$'s candidate will win the final election.\footnote{``Do it. I will personally write you a campaign cheque now, on behalf of this country, which does not want you to be president, but which badly wants you to run.''\\ --John Oliver, on the prospect of Trump running in 2016.} The authors model the situation in a one-dimensional Hotelling framework, and consider three scenarios: all voters are sincere, only voters of party $B$ are strategic, or all voters are strategic. They find that the worst outcome in terms of social welfare is when only voters of party $B$ are strategic, but the best outcome is when all voters strategise -- the effect is explained by the fact that the voters of party $A$ are motivated to vote for a more moderate candidate to counter the behaviour of the voters of party $B$.

In a series of papers, Lehtinen uses numeric experiments to assess the welfare effects of strategic voting under sequential majority \cite{Lehtinen2007Agendas}, Borda \cite{Lehtinen2007Borda}, approval voting, and plurality \cite{LehtinenApproval}. In his framework voters have cardinal utility over three candidates, and receive a noisy signal about the support the other candidates have. Each voter assumes the other voters are acting sincerely, and then votes in accordance with expected utility maximisation. The result is that strategic voting allows voters to express intensity of preference -- a voter for whom candidate $a$ is only slightly better than $b$ is less likely to manipulate in favour of $a$ than a voter for whom $a$ is much better. He finds that under sequential majority, Borda, and plurality utilitarian efficiency (the probability of electing the candidate maximising social welfare) increases under strategic behaviour, particularly if the correlation between the utilities of the different candidates is high. Under plurality the effect is particularly marked, the utilitarian efficiency increasing from $\approx 35\%$ to as much as $95\%$. The lower the correlation, the less pronounced the effect, and under approval voting there is very little difference between the sincere and strategic scenarios. In a second series of experiments the Condorcet efficiency of the rules is compared; here strategic behaviour is harmful.

Lu et al. \cite{Lu2012} consider a setting where a group of manipulators want to elect a target candidate, but possess only partial, probabilistic knowledge of the preferences of the sincere voters. The main focus of their work is on how to compute the optimum strategy of the manipulators, but they also assess the damage the manipulators can do to social welfare, i.e.\ the difference in the utility derived by the sincere voters before and after manipulation. The authors train a Mallows model on Dublin West electoral data and simulate voter behaviour under the Borda rule. They find that while the probability of the manipulators being able to influence the election can be very high, the damage to social welfare is low -- never more than $5\%$. This is explained by the fact that a manipulation is more likely to be successful in favour of a candidate that already enjoys broad support among the sincere voters, rather than a candidate who is reviled by everyone.

Bassi \cite{bassi2015voting} studied the strategic behaviour of human voters in laboratory experiments. Every game in the experiment consisted of five voters and four candidates, using one of plurality, Borda, or approval voting. The main focus of the paper was on whether humans vote as predicted by the iterated elimination of dominated strategies, thus every profile used had a dominance solution. The welfare of the outcomes was measured in terms of Condorcet efficiency and social welfare (the sum of the participants' monetary payoffs), and compared to the theoretical prediction for sincere agents and for strategic agents playing the equilibrium solution. Plurality voting was found to outperform Borda and approval voting in terms of social welfare, but lost in terms of Condorcet efficiency.

\subsection{Our contribution}

We perform numeric experiments to evaluate the welfare effects of strategic voting by a small panel of voters. Our experiments cover 15 scoring rules, 3 measures of welfare, and 9 statistical cultures. Among our findings, we find that concave rules suffer the most welfare loss under manipulation, to the point that if our goal is to maximise Rawls or Nash welfare, it is rarely a good idea to use a rule that na\"ively maximises said Rawls or Nash welfare; that convex rules are resistant to manipulation, which makes them more likely to elect the best outcome under a Mallows model; and that $(m/2)$-approval performs exceedingly well under Euclidean preferences, for all the welfare measures studied.

\section{Preliminaries}

\subsection{Voting concepts}

Let $\CV$, $|\CV|=n$, be a set of voters, $\CC$, $|\CC|=m$, a set of candidates,  and $\CL(\CC)$ the set of linear orders over $\CC$. Every voter is associated with some $\succeq_i\in\CL(\CC)$, which denotes the voter's preferences. A profile $P\in\CL(\CC)^n$ is an $n$-tuple of preferences, $P_i$ is the $i$th component of $P$ (the preferences of voter $i$), and $P_{-i}$ the preferences of all the other voters.

A voting rule is a mapping:
$$f:\CL(\CC)^n\ria \CC,$$
I.e., it is a rule which associates each profile with a candidate, who is the \emph{election outcome}.

We are interested in a class of voting rules known as scoring rules.
\begin{definition}
A \emph{scoring rule} is a voting rule defined by a sequence of scores, $s_1,\dots,s_m$ satisfying $s_i\geq s_{i+1}$ and $s_1>s_m$. Using $\text{pos}(i,c)$ to denote the position of candidate $c$ in voter $i$'s ballot, the score of $c$ is:
$$\text{score}(c)=\sum_{i\leq n} s_{\text{pos}(i,c)}.$$
The candidate with the highest score is the election outcome, ties are broken lexicographically.

In this paper we are interested in the following scoring rules:
\begin{itemize}
    \item \emph{$k$-approval} is the scoring rule with $s_i=1$ for $i\leq k$, and $s_j=0$ for $j>k$. I.e., the best $k$ candidates get one point, the rest 0. 1-approval is also known as \emph{plurality}.
    \item \emph{$k$-Borda} is the scoring rule with $s_i=\max(k-i+1,0)$. I.e., the top candidate gets $k$ points, then $k-1$, $k-2$, and so on. $(m-1)$-Borda is also known as \emph{Borda}.
    \item A \emph{geometric scoring rule with parameter $p$} \cite{kondratev2019should}, or \emph{geometric $p$} for short, is the scoring rule with $s_i=p^{m-i}$ if $p>1$ (the \emph{convex} rules) and $s_i=1-p^{m-i}$ if $p<0$ (the \emph{concave} rules). For example, with 5 candidates and $p=2$ the scoring vector is $16,8,4,2,1$.
    \item The \emph{Nash rule} is the scoring rule with $s_i=\log(m-i)$ for $i<m$, and $s_m=-n\log(m-1)$.
\end{itemize}
\end{definition}

We will evaluate the election outcomes with three ordinal welfare functions, normalised to give an outcome between 0 and 100.

\begin{definition}
Suppose $c$ is the election outcome. The \emph{Borda welfare} of the outcome is the sum of the Borda scores of $c$, normalised by the hypothetical Borda score of a candidate ranked first by all:
$$\text{Borda}(c)=100\frac{\sum_{i\leq n} (m-\text{pos}(i,c))}{(m-1)n}.$$
The \emph{Rawls welfare} of the outcome is the Borda score given to $c$ by the voter that ranks $c$ the lowest, normalised by $(m-1)$:
$$\text{Rawls}(c)=100\frac{\min_{i\leq n} (m-\text{pos}(i,c))}{(m-1)}.$$
The \emph{Nash welfare} of the outcome is the product of the Borda scores of $c$, normalised by taking the $n$th root:
$$\text{Nash}(c)=100\frac{\sqrt[n]{\prod_{i\leq n} (m-\text{pos}(i,c))}}{(m-1)}.$$
\end{definition}

Borda welfare is based on the utilitarian principle of choosing a candidate with the best average rank. Rawls welfare is based on the egalitarian notion of choosing a candidate that will minimise the misery of the unhappiest voter. Nash welfare, while originally proposed as a solution to a bargaining problem, is often proposed as a means to achieve a balance of the utilitarian and egalitarian principles \cite{moulin2004fair,MasthoffGroupRecommendations}. It should be noted that, unlike Borda and Rawls welfare, the Nash-maximising outcome depends on the choice of the zero point. In this paper, we value the last position at 0, but the performance of the voting rules with respect to Nash welfare would have differed somewhat with a different choice.

Under sincere behaviour, the Borda rule, by definition, maximises Borda welfare.

Rawls welfare is maximised by generalised antiplurality, which is not a scoring rule per se, but a \emph{generalised scoring rule} \cite{Smith1973} -- candidates are first ranked by their $(m-1)$-approval score, first-order ties are broken by the $(m-2)$-approval score, second-order ties by the $(m-3)$-approval score, and so on. We do not investigate generalised antiplurality directly, but the rule is equivalent to a geometric scoring rule with a sufficiently small $p$ \cite{kondratev2019should}; thus the geometric rule with $p=0.5$ is a close proxy to the Rawlsian optimum.

Given that $\log(xy)=\log(x)+\log(y)$, a candidate maximises Nash welfare if and only if it maximises the sum of the logarithms of its Borda points; this is the motivation behind the Nash rule. Unfortunately this does not give us a scoring rule because $\log(0)$ is undefined, which is why we set $s_m=-n\log(m-1)$; thus, no amount of first places will compensate for a single last place.

The main interest of this paper is measuring the effect of strategic voting on these measures of welfare.

\begin{definition}
Consider a voting rule $f$ and a profile $P$. A \emph{strategic vote} for voter $i$ is a $P_i'$ such that:
$$f(P_i',P_{-i})\succ_i f(P_i,P_{-i}).$$
In other words, if voter $i$ casts $P_i'$ instead of $P_i$, then the election outcome is preferable for voter $i$.

A voter may have many strategic votes available to him. We thus define his \emph{optimal strategy} to be a $P_i^*$ such that:
$$P_i^*\in\argmax_{P_i'} f(P_i',P_{-i}),$$
the $\argmax$ operator is understood with respect to voter $i$'s preferences over the election outcomes.
\end{definition}

Against this standard definition of manipulation two criticisms are often levied: 1) the probability of a single voter affecting the outcome of an election is negligibly small, and 2) it is unreasonable to suppose this voter has access to information about the other voters' preferences, which he needs to compute his strategic vote. In the political interpretation of voting, both arguments are undoubtedly valid, and a reasonable model should account for strategic voting by groups and incomplete information, such as the model of \cite{Lu2012} or \cite{Lehtinen2007Borda}. However, if we consider the interpretation of a small group of experts making a choice between candidates with the aim of identifying the one which is ``best'', then the standard definition is reasonable enough -- it would not take a great deal of social engineering to find out how one's colleagues plan to rank interview candidates (say, ask them during a coffee break), and a single unethical judge can be enough to skew the outcome of a sporting context. Since our simulations will focus on the case where the number of voters is small and we are measuring the outcome in terms of numerical measures of welfare, we believe the standard definition of manipulation is sufficient.

\subsection{Statistical cultures}

We will generate voter profiles from four theoretical families. In each case, the voters are sampled i.i.d.

\begin{itemize}
    \item Impartial culture: given $n$ voters and $m$ candidates, each voter's preferences are drawn uniformly at random  from all $m!$ possible preference orders.
    \item $k$-Euclidean: every voter and candidate is generated, uniformly at random, in $[0,1]^k$, the $k$-dimensional unit cube. A voter's preferences are determined by Euclidean distance to the candidates -- the closer the candidate, the more preferred.
    \item Mallows model: given a dispersion parameter $\phi$, $0<\phi<1$, and a reference order $\sigma$, a voter is assigned a preference order $r$ with probably $\frac{1}{Z}\phi^{d(\sigma, r)}$. $d(\sigma,r)$ is the Kendall-tau distance between $\sigma$ and $r$, and $Z$ is a normalisation constant to make sure the probabilities add to one.  
    \item Mixed Mallows model: given dispersion parameters $\phi_1,\dots,\phi_k$, reference orders $\sigma_1,\dots,\sigma_k$, and probabilities $p_1,\dots,p_k$, a voter is assigned a preference order $r$ with probably $\sum_{i\leq k}p_i\frac{1}{Z_i}\phi_i^{d(\sigma_i, r)}$. That is, each $(\phi_i,\sigma_i)$ defines a Mallows model, and we choose which Mallows model to sample from with the probabilities $p_1,\dots, p_k$.
\end{itemize}
It should be noted that a mixed Mallows model will behave very differently depending on what reference orders are used, e.g.\ drawing profiles from a mixture of $a\succ b\succ c\succ d$ and $b\succ a\succ c\succ d$ will result in much more correlated preferences than a mixture of $a\succ b\succ c\succ d$ and $d\succ c\succ b\succ a$. In this paper, when we sample from the mixed Mallows model, it is understood that the reference orders are chosen randomly for each profile sampled.

In addition to these theoretical cultures we use two cultures based on empirical data from Preflib \cite{Mattei2013PrefLibAL}. The first, Mallows sushi, is a mixed Mallows model trained by Lu and Boutilier \cite{Lu2014} on a dataset of 5,000 preferences over 10 sushi types (\url{https://www.preflib.org/data/ED/00014}).

The second is based on skating data, consisting of judges' rankings of athletes in various events. The preferences in this data are highly correlated, and in most events the winner is unanimous. To deal with this we hand-picked an event with the most disagreement among the judges (\url{https://www.preflib.org/data/ED/00006/00000046}), and used a simple bag-of-preferences model (sampling the judges' rankings with replacement).

\subsection{Experimental setup}

The simulations are based on varying the following parameters.

 Voting rules:
\begin{itemize}
    \item Borda family ($m-1$, $m/2$, $m/4$, $5$).
    \item Approval family ($m/2$, $m/4$, $5$, $1$).
    \item Geometric family ($2$, $1.5$, $1.2$, $0.8$, $0.65$, $0.5$).
    \item Nash rule.
\end{itemize}

Theoretical cultures:
\begin{itemize}
    \item Impartial culture.
    \item Euclidean family (1, 2, 5 dimensions).
    \item Mallows family (dispersion parameters 0.8, 0.5).
    \item Mixed Mallows (two equiprobable components, dispersion parameter 0.5).
\end{itemize}

Welfare measure:
\begin{itemize}
    \item Borda welfare.
    \item Rawls welfare.
    \item Nash welfare.
\end{itemize}

Voter behaviour:
\begin{itemize}
    \item All voters sincere.
    \item Voter 1 acts optimally.
\end{itemize}

The procedure for generating welfare results is as follows:
\begin{enumerate}
    \item Fix $n=10$. For each choice of $m\in\set{3,\dots,100}$, voting rule, theoretical culture, behaviour, and welfare measure generate 10,000 profiles, measure the welfare of the election outcome, and return the average.
    \item Fix $m=10$. For each choice of $n\in\set{3,\dots,100}$, voting rule, behaviour, and welfare measure generate 10,000 profiles from the Mallows sushi culture, measure the welfare of the election outcome, and return the average.
    \item Fix $m=30$. For each choice of $n\in\set{3,\dots,100}$, voting rule, behaviour, and welfare measure generate 10,000 profiles from the skating bag culture, measure the welfare of the election outcome, and return the average.
\end{enumerate}

\section{Results}

\begin{figure}
    \centering
    \begin{tabular}{cc}
           \subcaptionbox{Borda welfare for IC, $n=10$.\label{fig:ICBorda}}{\includegraphics[width = 6cm]{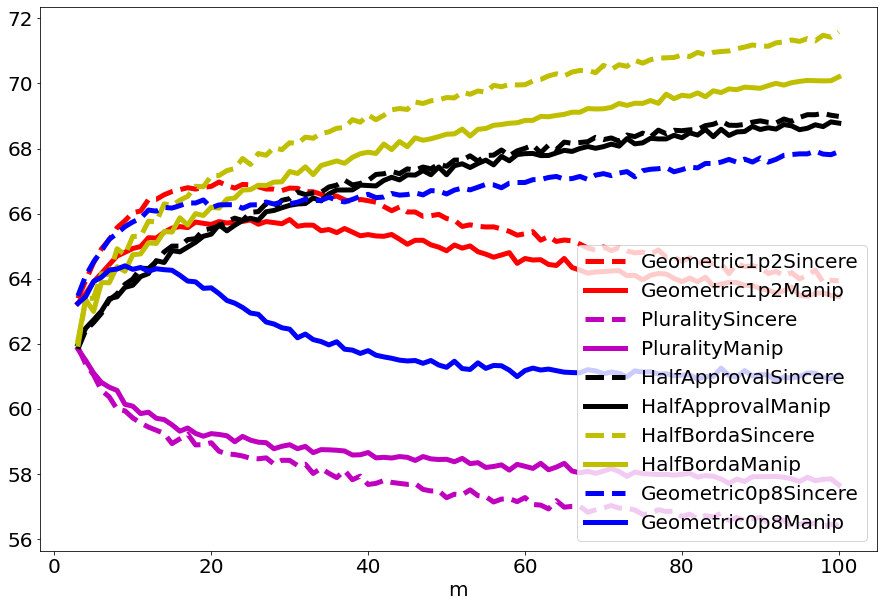}}
           &\subcaptionbox{Nash welfare for IC, $n=10$.\label{fig:ICNash}}{\includegraphics[width = 6cm]{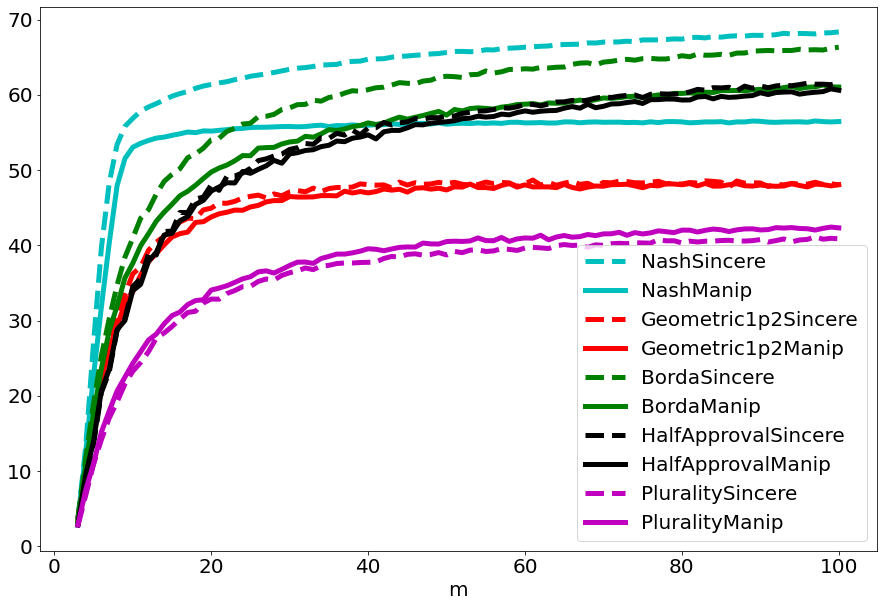}}\\
         \subcaptionbox{Borda welfare for 2-Euclidean, $n=10$.\label{fig:2DBorda}}{\includegraphics[width = 6cm]{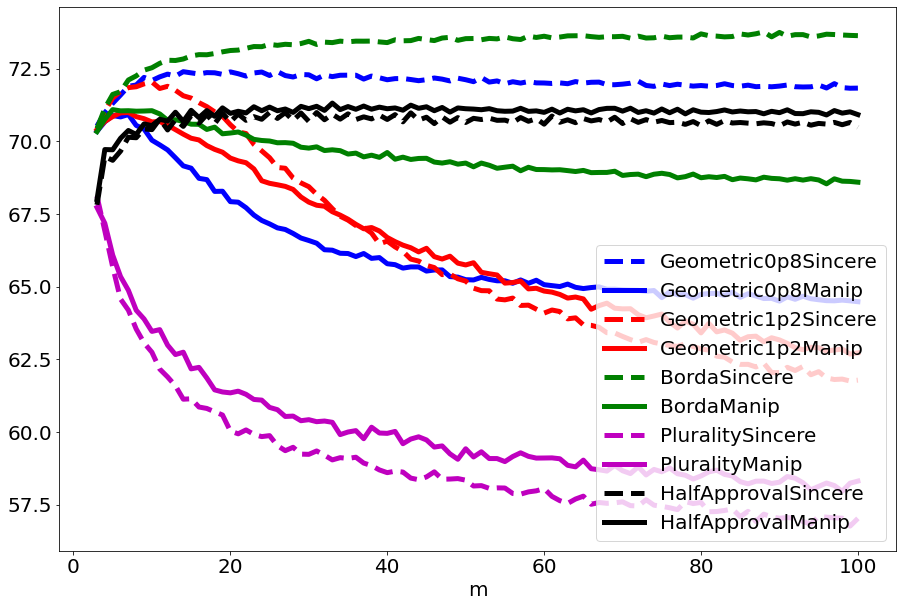}}& 
         \subcaptionbox{Rawls welfare for 5-Euclidean, $n=10$.\label{fig:5DRawls}}{\includegraphics[width = 6cm]{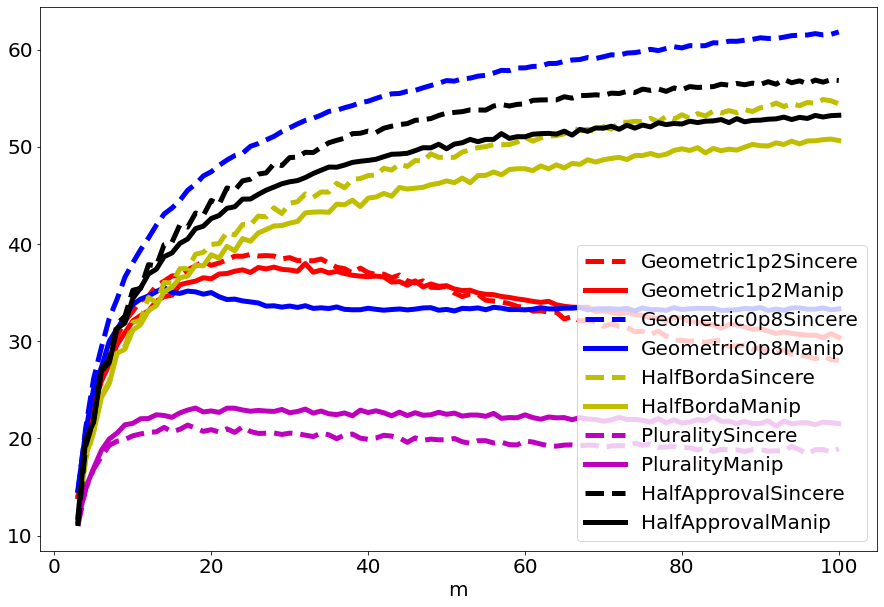}}\\
         \subcaptionbox{Rawls welfare for Mallows $0.8$, $n=10$.\label{fig:mallows8}}{\includegraphics[width = 6cm]{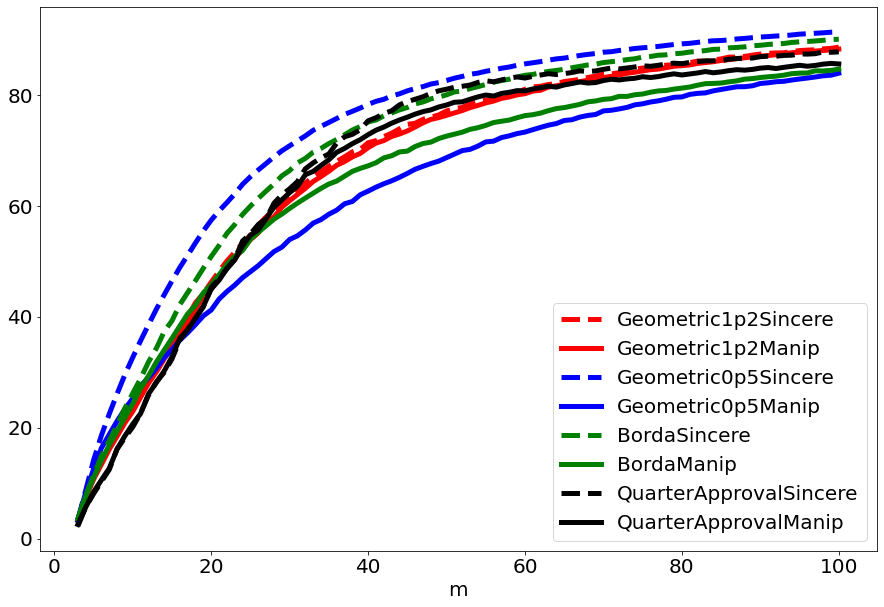}}& 
         \subcaptionbox{Nash welfare for Mallows $0.5$, $n=10$.\label{fig:mallows5}}{\includegraphics[width = 6cm]{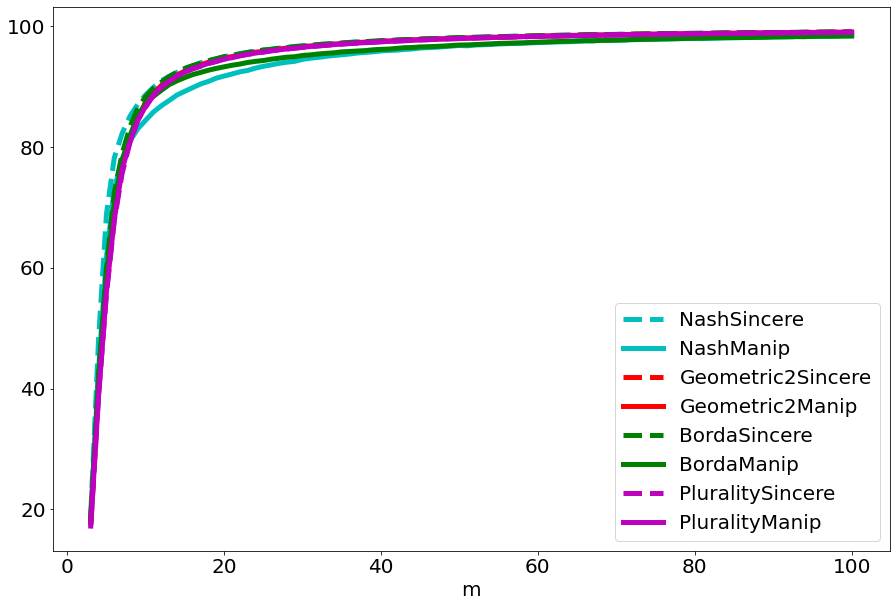}}\\
         \subcaptionbox{Borda welfare for sushi, $m=10$.\label{fig:sushi}}{\includegraphics[width = 6cm]{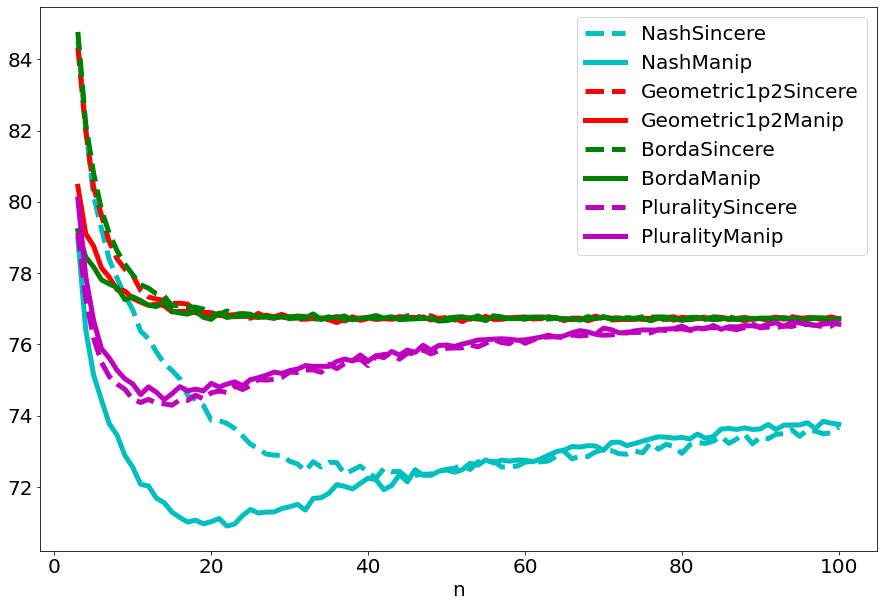}}& 
         \subcaptionbox{ Borda welfare for skating bag, $m=30$.\label{fig:skating}}{\includegraphics[width = 6cm]{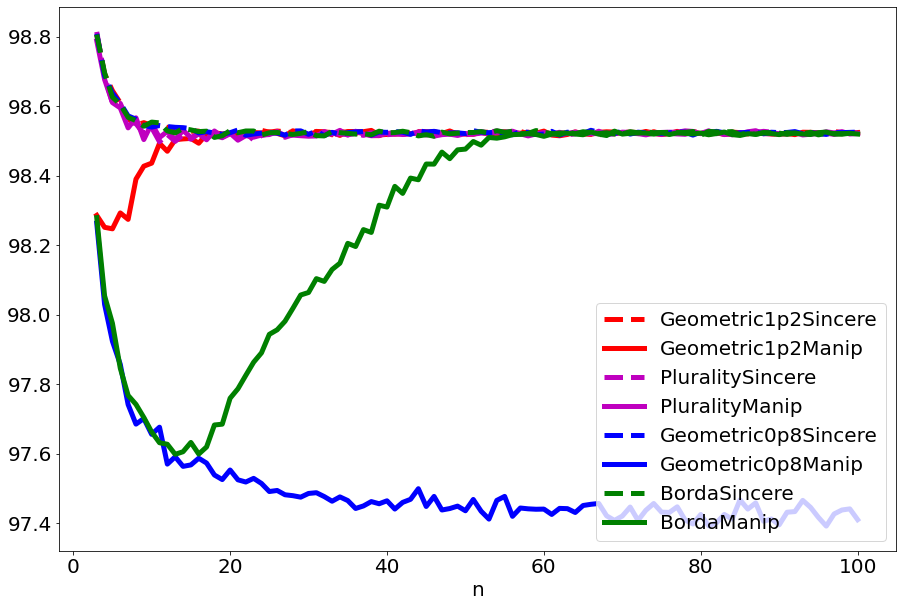}}
    \end{tabular}
    \caption{Selected plots of welfare vs number of voters/candidates.}
\end{figure}

In this section we discuss the most interesting or representative results. The results for all the studied scenarios, along with the data, code, and instructions for replication, can be found on the Github.\footnote{\url{https://github.com/EIanovski/WelfareManipulation}.}

\subsection{Impartial culture}

Impartial culture is often seen as a worst-case for manipulation, since it increases the odds of a close election where strategic voting can make a difference. In terms of consequences on welfare, however, this turns out to be rather benign; though most scoring rules lose welfare after manipulation, Borda (and its truncated version) remains the best choice for maximising Borda welfare (\autoref{fig:ICBorda}), Geometric 0.5 for Rawls welfare, and Nash for Nash (at least for $m<40$, after which it is overtaken by Borda. See \autoref{fig:ICNash}).

Nevertheless, the results with impartial culture allow us to make some general observations about the behaviour of scoring rules, which remain true in subsequent experiments.

Under sincere behaviour, concave rules expectedly do well in terms of Rawls and Nash welfare, while the Nash rule performs well in all three. However, these rules are very susceptible to manipulation, and experience the largest welfare drops. In the case of impartial culture, these drops are insufficient to dethrone the geometric rule with $p=0.5$ as the best choice for maximising Rawls welfare, but when we move on to other cultures we shall see the concave rules perform very poorly.

Convex geometric rules prove to be resistant to manipulation, and experience very minor welfare drops under strategic behaviour. However, their overall welfare properties are poor, so their stability does little to recommend them. Plurality is the most convex rule imaginable, and it is one of the few rules to consistently yield \emph{higher} welfare in the presence of manipulation. However, even with the gain from manipulation, it consistently ranks as the worst rule in terms of welfare.

Borda lies on the border of convex and concave rules \cite{kondratev2019should}, and its properties lie between the two: loss of welfare after manipulation is tangible, but nowhere near the extent of concave rules. Though rarely the best rule in terms of welfare under manipulation, it routinely does pretty well.

5-approval and 5-Borda fare poorly, tending towards the plurality outcome as the number of candidates grows. This is unfortunate since the appeal of these rules is in their low cognitive burden on the voter, requiring the voter to rank 5 candidates rather than $m$; but it seems that if we value welfare, we rarely can get by without asking the voters for their full rankings. $(m/2)$-approval and $(m/2)$-Borda fare much better; $(m/2)$-Borda often outperforms full Borda, while $(m/2)$-approval is the clear winner under Euclidean preferences, to which we turn next.

\subsection{Euclidean cultures}

The key feature of the Euclidean cultures is the existence of a centre -- candidates close to the centre of the unit cube are reasonably close to all voters, and thus tend to have high Borda, Rawls, and Nash welfare. This goes a long way to explain the outstanding performance of $(m/2)$-approval in this setting (\autoref{fig:2DBorda}, \autoref{fig:5DRawls}) -- the centre candidates are likely to be in the top half of most voters' ballots, and thus accumulate too many points for the manipulator to do much beyond choosing which centre candidate in particular should win.

As with impartial culture, the welfare of concave rules drops sharply under manipulation, but this time the effect is that these rules never maximise their respective welfare. Under 2- and 5-Euclidean, indeed, the Borda rule is the best choice for Rawls and Nash welfare after $(m/2)$-approval.

The dimension of the space seems to make a difference. Under 5-Euclidean the Borda rule performs a lot better, with $(m/2)$-Borda yielding the highest Borda welfare and Nash welfare for $m>40$ (\autoref{tab:best}). Interestingly, $(m/4)$-approval also outperforms $(m/2)$-approval in terms of Borda welfare.

\subsection{Mallows models}

The Mallows model models the situation where there is an underlying objective truth, and voter preferences are noisy signals of this true order. This noise is modelled as swaps of neighbouring candidates in the order (Kendall-tau distance), so the less swaps separate an order from the objective truth, the more likely it is to be generated. The result of this is that voter preferences are highly correlated, and as the number of candidates increases, so does the probability of the outcome being unanimous -- with 100 candidates, the probability of a swap occurring at the very top of a voter's preferences is very low. The dispersion parameter affects how soon we reach the point where all voters are likely to agree on the top candidates (\autoref{fig:mallows8} vs \autoref{fig:mallows5}).

This is where convex rules come to the fore. The extra weight they give to the top candidates allows the candidates ranked high in the objective order to accumulate an overwhelming lead, reducing the manipulator's ability to harm social welfare. The result is that these rules are optimal at once with respect to Borda, Rawls, and Nash welfare. Concave rules, including Nash and Borda, give the manipulator enough power to bury a candidate ranked highly by the others, which harms their welfare for small values of $m$. However as we have observed, as $m$ increases we head towards consensus, where the manipulator will have no incentive to manipulate.

The situation is different in our mixed Mallows model, which we remind the reader consists of two equiprobable components with dispersion parameters 0.5; the reference orders are sampled randomly for each profile. Convex rules again display poor welfare properties overall, while Nash and the concave geometric rules lose a lot from manipulation. The winners are members of the Borda and approval families, with $(m/2)$-Borda and $(m/4)$-approval performing well in terms of all welfare measures.

\subsection{Mallows sushi}

As another Mallows mixture, one might expect the Mallows sushi model to perform similar to our mixed Mallows model, but this is not the case. It appears that there is such a thing as the objectively best sushi (fatty tuna, for the curious), and as the number of voters increases, all voting rules converge to this result (\autoref{fig:sushi}). Borda, the convex rules, and the truncated rules converge rapidly; plurality and the highly concave rules slowly; Nash the slowest of all. Manipulation is only really tangible for the highly concave rules and Nash, as it gives voters the ability to effectively veto a choice of sushi.

\subsection{Skating bag}

The skating bag culture is an example of a situation where there is a clear ``best'' candidate, who maximises at once Borda, Rawls, and Nash welfare, and all reasonable voting rules, plurality included, elect this best candidate. The damage of manipulation, therefore, measures the ability of a single voter to force through his preferred outcome in spite of overwhelming social consensus. A scoring rule needs to be top-heavy to resist this behaviour, so the high positions of the best candidate in the sincere voters' ballots will outweigh any shenanigans by the manipulator. As plurality is the top-heavy rule par excellence, this is one of the rare situations where this rule shines (\autoref{fig:skating}). Convex geometric rules are vulnerable when the number of voters is small, but from about 10 voters onwards the sincere votes begin to outweigh the manipulator's endeavours. The Borda rule is well-known for the scope it gives a manipulator to cause mischief, and it is only after we have 50 voters that the threat under Borda is liquidated.

\renewcommand{\arraystretch}{1.3}
\begin{table}
    \centering
    \begin{tabular}{@{}p{3cm}p{3cm}p{3cm}p{3cm}@{}}
    \toprule
         Culture&  Borda welfare&Rawls welfare&Nash welfare\\
         \midrule
         Impartial culture& 
         \makecell[tl]{\textbf{Borda} ($m<25$),\\ $(m/2)$-Borda\\\quad($m>25$).}&
         \makecell[tl]{\textbf{Geometric 0.5},\\ $(m/2)$-approval\\\quad($m=100$).}&
         \makecell[tl]{Borda ($m>40$),\\ \textbf{Nash} ($m<40$),\\ \textit{($m/2$)-approval}\\\quad\textit{$(m>40).$}}\\
         1-Euclidean&
         \makecell[tl]{\textbf{Borda} ($m<12$),\\ ($m/2$)-approval\\\quad($m>12$).}&
         \makecell[tl]{($m/2$)-approval. 
         }
         &
         \makecell[tl]{
         ($m/2$)-approval. 
         }\\
         2-Euclidean&
         \makecell[tl]{\textbf{Borda} ($m<12$),\\ ($m/2$)-approval\\\quad($m>12$).}&
         ($m/2$)-approval.&
         \makecell[tl]{
         ($m/2$)-approval. 
         }\\
         5-Euclidean& 
         \makecell[tl]{ $(m/2)$-Borda, \\ \textit{($m/4$)-approval}\\\quad\textit{($m>95$).}}&
         \makecell[tl]{
         ($m/2$)-approval. 
         }
         &
         \makecell[tl]{
         ($m/2$)-approval\\\quad($m<40$),
         \\$(m/2)$-Borda\\\quad($m>40$).}\\
         Mallows 0.8& 
         Geometric 1.2.&
         \makecell[tl]{Geometric 1.2\\\quad($m>65$),\\ ($m/4$)-approval\\\quad($25<m<65$),\\ Borda ($m<25$).}&
         \makecell[tl]{Geometric 1.2\\\quad($m>15$),\\ \textbf{Nash} ($m<15$).}\\
         Mallows 0.5& 
         \makecell[tl]{Geometric $p>1$,\\ Plurality.}&
         5-approval.&
         \makecell[tl]{Geometric 2,\\ \textit{Plurality.}}\\
         Mixed Mallows& 
         \makecell[tl]{$(m/2)$-Borda,\\ \textit{($m/4$)-approval}\\\quad\textit{($m>30$).}}&
         ($m/4$)-approval.&
         \makecell[tl]{($m/4$)-approval\\\quad($m>20$),\\ ($m/2$)-Borda\\\quad($m>80$),\\ 5-approval ($m<20$).}\\
         \bottomrule
    \end{tabular}
    \medskip
    \caption{Best scoring rules post manipulation with $n=10$, $m>10$. Boldface denotes rules that maximise welfare under sincere behaviour. Italics denote rules that are almost as good as the best, but much simpler.}
    \label{tab:best}
\end{table}

As for concave rules, the threat never goes away. Even in an electorate of 100 voters, a single manipulator can force through a socially suboptimal outcome. A single last position does too much damage to the best candidate, giving the manipulator effective veto power over the outcome.

\section{Conclusions}

We summarise the three main takeaways of this study:

\begin{itemize}
    \item Manipulation makes a difference. With the exception of impartial culture, or profiles with a small number of candidates, the welfare-maximising rule with a manipulator is never the same as under sincere behaviour (\autoref{tab:best}). The effect of manipulation in our framework is almost always negative.
    \item Top-heavy rules such as convex geometric rules and plurality are resistant to manipulation, losing little, or even gaining welfare in the presence of a manipulator. However, their welfare properties are so poor that this property does little to recommend them. The exception is in the case of highly correlated cultures such as Mallows models, where the additional weight these rules gives to the top candidate stymies attempts at manipulation.
    \item Concave geometric rules and the Nash rule are very susceptible to manipulation, to the point that Borda or even a convex rule is often better at maximising Rawls/Nash welfare than the rules designed for that purpose. This is an issue because empirical evidence suggests that humans value a mixture of the egalitarian and utilitarian principles, and given a choice will choose a voting rule that strikes a balance between the two \cite{Ambuehl2021,Frohlich1987,MasthoffGroupRecommendations}. If these rules fail to deliver in the face of strategic behaviour, the question must be posed: what voting rule should a society choose, if it seeks to strike a balance between utilitarian and egalitarian principles?
\end{itemize}

\bibliographystyle{splncs04}
\bibliography{references}
\end{document}